\documentclass[pra,showpacs,twocolumn,floatfix,groupaddress,superscriptaddress,footinbib]{revtex4}
\usepackage{times,amsmath,amsfonts,amssymb,latexsym}
\usepackage{graphicx,epsf}
\usepackage{subfigure}
\newcommand{\be}{\begin{equation}}
\newcommand{\ee}{\end{equation}}
\newcommand{\ba}{\begin{eqnarray}}
\newcommand{\ea}{\end{eqnarray}}

\newcommand{\ignore}[1]{}

\newcommand{\partder}[2]{\frac{\partial #1}{\partial #2}}
\newcommand{\ket}[1]{\left | {#1} \right \rangle }

\newcommand{\bra}[1]{\left \langle {#1} \right | }

\newcommand{\hadj}[1]{{#1}^{\dagger}}
\newcommand{\conj}[1]{{#1}^{\ast}}
\newcommand{\cosine}[1]{\mathrm{cos}\left ( {#1}\right )}
\newcommand{\sine}[1]{\mathrm{sin}\left ( {#1}\right )}
\newcommand{\cosinep}[2]{\mathrm{cos}^{#2}\left ( {#1}\right )}
\newcommand{\sinep}[2]{\mathrm{sin}^{#2}\left ( {#1}\right )}

\newcommand{\trace}[1]{\mathrm{Tr}\left ( {#1}\right )}
\newcommand{\parent}[1]{\left( {#1} \right)}

\newcommand{\absval}[1]{\left| {#1} \right|}
\newcommand{\sset}[1]{\left\lbrace {#1} \right\rbrace }

\def\CC{{\rm\kern.24em \vrule width.04em height1.46ex depth-.07ex
    \kern-.30em C}}
\def\P{{\rm I\kern-.25em P}}
\def\RR{{\rm
         \vrule width.04em height1.58ex depth-.0ex
         \kern-.04em R}}

\def\bbbc{{\mathchoice {\setbox0=\hbox{$\displaystyle\rm C$}\hbox{\hbox
to0pt{\kern0.4\wd0\vrule height0.9\ht0\hss}\box0}}
{\setbox0=\hbox{$\textstyle\rm C$}\hbox{\hbox
to0pt{\kern0.4\wd0\vrule height0.9\ht0\hss}\box0}}
{\setbox0=\hbox{$\scriptstyle\rm C$}\hbox{\hbox
to0pt{\kern0.4\wd0\vrule height0.9\ht0\hss}\box0}}
{\setbox0=\hbox{$\scriptscriptstyle\rm C$}\hbox{\hbox
to0pt{\kern0.4\wd0\vrule height0.9\ht0\hss}\box0}}}}
\def\bbbz{{\mathchoice {\hbox{$\sf\textstyle Z\kern-0.4em Z$}}
{\hbox{$\sf\textstyle Z\kern-0.4em Z$}}
{\hbox{$\sf\scriptstyle Z\kern-0.3em Z$}}
{\hbox{$\sf\scriptscriptstyle Z\kern-0.2em Z$}}}}

\newcommand{\PRL}{Phys.~Rev.~Lett.}
\newcommand{\PRA}{Phys.~Rev.~A}

\usepackage{hyperref}

\begin{document}

\title{Universality and robustness of revivals in the transverse field XY model}
\author{Juho H\"{a}pp\"{o}l\"{a}}
\affiliation{Aalto University, Helsinki, Finland}
\affiliation{University of Helsinki, Finland}
\affiliation{Perimeter
Institute for Theoretical Physics, 31 Caroline St. N, N2L 2Y5,
Waterloo ON, Canada}
\author{G\'{a}bor B. Hal\'{a}sz}
\affiliation{Perimeter Institute for Theoretical Physics, 31
Caroline St. N, N2L 2Y5, Waterloo ON, Canada}
\affiliation{Trinity
College, University of Cambridge, Trinity Street, Cambridge CB2 1TQ,
UK}
\author{Alioscia Hamma}
\affiliation{Perimeter Institute for Theoretical Physics, 31
Caroline St. N, N2L 2Y5, Waterloo ON, Canada}

\begin{abstract}

We study the structure of the revivals in an integrable quantum
many-body system, the transverse field XY spin chain, after a
quantum quench. The time evolutions of the Loschmidt echo, the
magnetization, and the single spin entanglement entropy are
calculated. We find that the revival times for all of these
observables are given by integer multiples of $T_{\mathrm{rev}}
\simeq L / v_{\mathrm{max}}$ where $L$ is the linear size of the
system and $v_{\mathrm{max}}$ is the maximal group velocity of
quasiparticles. This revival structure is universal in the sense
that it does not depend on the initial state and the size of the
quench. Applying non-integrable perturbations to the XY model, we
observe that the revivals are robust against such perturbations:
they are still visible at time scales much larger than the
quasiparticle lifetime. We therefore propose a generic connection
between the revival structure and the locality of the dynamics,
where the quasiparticle speed $v_{\mathrm{max}}$ generalizes into
the Lieb$-$Robinson speed $v_{LR}$.

\end{abstract}


\maketitle

\section{Introduction} \label{sec-intro}

The behavior of quantum many-body systems away from equilibrium has
recently become the object of intense experimental study. Ultra-cold
atoms in optical lattices feature both large phase coherence times
and a high degree of controllability, making it possible to observe
quantum coherent dynamics \cite{coldatoms}. The potential
technological implications are profound because the manipulation of
coherent quantum dynamics is at the root of the possibility of
building a quantum computer.

In terms of theoretical description, systems away from equilibrium
are considerably more complicated than their equilibrium
counterparts. Whereas equilibrium systems can be understood by means
of standard methods like mean field theory and renormalization
group, we lack analogous methods  for understanding non-equilibrium
physics \cite{review}. There are no obvious generalizations of such
standard methods at equilibrium, and in particular, it is not clear
to what extent the non-equilibrium dynamics of quantum systems
{features} universality. An important example is the
universality of the Kibble$-$Zurek mechanism to compute the density
of defects as the external temperature is tuned in time
\cite{kibble-zurek}.

Experiments on cold atomic gases give us the opportunity to observe
a genuine quantum evolution in a system that is very close to be
isolated. We can then {address} questions like {the
possibility} of equilibration in {a system with} unitary
evolution and the process of thermalization in a closed quantum
system. {These topics have recently} found a renewed interest
{along} with other fundamental problems in quantum
statistical mechanics \cite{tasaki, winter}. In the case of such
systems, {the} non-equilibrium dynamics is obtained
{by making some parameters of the system Hamiltonian
time-dependent and hence casting the system off equilibrium}. The
time dependence of the Hamiltonian can be adiabatically slow
\cite{polkovnikov} or it can change abruptly \cite{quench1, quench2}
in which case the process is called a \emph{quantum quench}. Among
other applications, the paradigm of the quantum quench has been
recently used to study the behavior of topological phases away from
equilibrium \cite{topquench}.

The understanding of non-equilibrium dynamics of isolated
interacting quantum systems is of fundamental importance to
understand quantum equilibration, a topic that has lately
experienced a new renaissance \cite{tasaki, winter, rigol1}. Due to
unitary time evolution, a \emph{finite} quantum system with a
non-trivial initial state can not converge to a steady state.
However, it has been shown in some remarkable papers \cite{cramer1,
cramer2} that thermalization of a finite subsystem is possible in
the infinite size limit. The system can be coarse grained by
choosing a partial set of local or macroscopic observables, and the
expectation values of these observables can in principle converge to
the ones in thermal equilibrium. In particular, an integrable system
does not thermalize: even local or macroscopic observables undergo
oscillations, or at best relax to equilibrium values that are
generally not the same as those predicted by the microcanonical
ensemble \cite{rigol2, rigol3}. Even for a non-integrable system,
thermalization does not always occur: in some cases there is only
relaxation to a non-thermal state depending on the initial
conditions \cite{mueller}.

The equilibration process has several characteristic time scales.
The largest one is the recurrence time $T_{\mathrm{rec}}$ at which
the system gets infinitesimally close to its initial state. It only
exists for finite systems, and diverges at least exponentially as a
function of the system size. The smallest time scale is the
relaxation time $T_{\mathrm{rel}}$ at which a given observable
relaxes to its long-term average value. Furthermore, there is a
third important time scale $T_{\mathrm{rev}}$ in between at which
\emph{revivals} occur: these are brief detachments from the average
value of an observable. Revivals typically last for a very short
time only (in comparison to the spacing between them), and their
magnitude decays in time as the equilibration process nears
completion.

In this paper, we investigate the structure of the revivals for an
integrable model, the transverse field XY spin chain. Our main
result is that this structure is universal in the sense that it does
not depend on the details of the quench and on the initial state.
{By applying} non-integrable perturbations to the system and
finding that the revival structure is surprisingly robust against
such perturbations, we argue that the revival structure is a
universal non-equilibrium property which follows from the locality
of the Hamiltonian.

\section{Formalism of the quantum quench} \label{sec-quench}

We consider a closed quantum system whose Hamiltonian $H(\lambda_1,
..,\lambda_R)$ depends on parameters  $\lambda_i$ representing the
coupling strengths of interactions and external fields. A quantum
quench is a sudden change in the Hamiltonian of the system. The
quantum system is originally prepared in the ground state $\rho_0$
of $H(\lambda^{(1)})$, and at time $t=0$ we switch the parameters to
different values $\lambda^{(2)}$. The system then evolves unitarily
with the quench Hamiltonian $H(\lambda^{(2)})$ according to $\rho(t)
= \mathcal U_t (\rho_0)$, where we define the superoperator
$\mathcal U_t (X) \equiv \exp(-i H t) X \exp(i H t)$. Unitary
evolution implies that a finite system can not converge to a steady
state $\overline{\rho}$, even weakly. The limit of $\rho(t)$ for
$t\rightarrow \infty$ does not exist unless the initial state
$\rho_0$ is trivial, e.g., an eigenstate. On the other hand, the
time average $\overline{\rho} = \lim_{t\rightarrow\infty}
t^{-1}\int^t_0 \rho(s) ds$ always exists and is given by the
$\rho_0$ totally dephased in the eigenbasis $\Pi_n = \ket{E_n}
\bra{E_n}$ of the Hamiltonian: $\overline{\rho}= \sum_n\Pi_n \rho_0
\Pi_n$. For a finite system, equilibration means that the
expectation values of macroscopic observables spend most of their
time very close to their average values.

An important quantity describing the time evolution is the Loschmidt
echo (LE) defined as $\mathcal L (t) = |\mbox{tr} (\exp(-i t
H)\rho_0)|^2$ which gives a measure of the {distance between
the} time evolved state $\rho(t)$ {and the} initial state
$\rho_0$. When the system undergoes a recurrence at $t =
T_{\mathrm{rec}}$ we have $\mathcal L (T_{\mathrm{rec}}) \simeq 1$.
The general expression for the LE can be written as
\begin{equation}
\mathcal L (t) = \sum_{n,m} p_n p_m e^{-i(E_n-E_m)t},
\label{eq-LE-1}
\end{equation}
where $p_n$ are the populations of the Hamiltonian eigenstates for
the initial state. It follows that the time average of the LE is
$\overline{\mathcal L} = \mbox{tr} (\overline{\rho}^2) = \sum_n
p^2_n$. The LE typically decays in a short time $T_{\mathrm{rel}}$
from $1$ to its average value $\overline{\mathcal L}$ around which
it oscillates. The relaxation time $T_{\mathrm{rel}}$ is $O(1)$ for
an off-critical quench, while it scales like $O(L^\zeta)$ with the
system size $L$ for a critical quench, that is, if $\lambda^{(2)}$
is a critical point, showing a critical \emph{slow-down} of the
system \cite{quan, venuti}.

Revivals are also visible in the LE as deviations from the average
value $\overline{\mathcal L}$. {We define revivals as time
instances at which the signal $\mathcal L (t)$ differs from
$\overline{\mathcal L}$ by more than three standard deviations.}
According to Eq. (\ref{eq-LE-1}), this happens when an {exceptionally}
large number of weights $p_n$ get partially back in phase. It is not
straightforward to understand directly from Eq. (\ref{eq-LE-1}) when
such a situation can occur in a generic quantum system, therefore we
consider the particular case of a simple one dimensional spin chain.

\section{Revivals in the XY model} \label{sec-xy}

\subsection{Exact solution by free fermions}

In this section, we consider an integrable (exactly solvable) model,
the one dimensional XY model of $N$ spins one half in a transverse
magnetic field. Since the model is exactly solvable, we can obtain
the whole spectrum and the eigenstate decomposition of the initial
state. {This leads to} an exact expression for the LE,
{and} the revival times can be extracted by inspecting its
time dependence.

The Hamiltonian of this spin chain is given by
\begin{equation}
H = - \frac{1}{2}\sum_{l=1}^{N} \parent{ \frac{1 + \eta}{2}
\sigma_l^x \sigma_{l+1}^x + \frac{1 - \eta}{2} \sigma_l^y
\sigma_{l+1}^y + h \sigma_l^z }, \label{eq-H-1}
\end{equation}
where $\eta$ is the anisotropy parameter, and $h$ is the external
transverse magnetic field. We assume cyclic boundary conditions
$\sigma_{N+1} = (-1)^{q} \sigma_1$. The periodic ($q=0$) and
antiperiodic ($q=1$) boundary conditions differ in $O(1/N)$ terms
and this difference usually does not affect the phase diagram or
other quantities in the thermodynamic limit. However, it can be
important in the LE that is typically exponentially small in $N$. In
the following, we shall see that the boundary conditions can have a
dramatic effect in the case of the critical quench. Note that the XY
model reduces to the quantum Ising model for $\eta = 1$, and to the
isotropic XX model for $\eta = 0$. The Hamiltonian exhibits two
regions of criticality: the XX model at $\eta = 0$ has a critical
region for $h \in (-1,1)$, while the XY regions of criticality are
the lines $h = \pm 1$.

The XY model can be diagonalized by a standard procedure
\cite{barouch}. In the first step, we map $\sigma_l$ to spinless
fermions by using a Jordan$-$Wigner transformation:
\begin{equation}
\sigma_l^z = 1 - 2\hadj{c}_l c_l, \quad \sigma_l^- =
\hadj{(\sigma_l^+)} = \hadj{c}_l e^{i \pi \sum_{j=1}^{l-1}
\hadj{c}_j c_j}. \label{eq-JW}
\end{equation}
The translational symmetry is then exploited by the Fourier
transform $c_{l} = \frac{1}{\sqrt{N}} \sum_{l=1}^{N} e^{ikl} c_k$,
where the momenta are quantized according to ${k}_n =\pi
(2n+1-q)/N$. Finally, after the Bogoliubov transformation $c_k =
\cos \theta_k \gamma_k + i \sin \theta_k \hadj{\gamma}_{-k}$, we
obtain a Hamiltonian describing non-interacting fermionic degrees of
freedom $\gamma_k$:
\begin{equation}
H = \sum_{k>0} \Lambda_{k} \parent{ \hadj{\gamma}_{k} \gamma_{k}+
\hadj{\gamma}_{-k} \gamma_{-k} - 1 }. \label{eq-H-2}
\end{equation}
The dispersion relation of these fermionic quasiparticles is given
by $\Lambda_k = \sqrt{\epsilon_k^2 + \eta^2 \sin^2 k}$ with
$\epsilon_k \equiv h - \cos k$, and the angle $\theta_k$ appearing
in the Bogoliubov transformation is $\theta_k = \tan^{-1}
\left[(\eta \sin k) / (\epsilon_k + \Lambda_k) \right]$.

\subsection{Loschmidt echo and revival times}

In a quantum quench, the system is prepared in the ground state of
$H(\lambda^{(1)})$, and then evolved with $H(\lambda^{(2)})$ at
$t>0$. It is useful to write the ground state $\ket{\psi
\parent{0}}$ of the initial Hamiltonian in terms of the eigenstates
of the quench Hamiltonian:
\begin{equation}
\ket{\psi \parent{0}} = \prod_{k>0} (\cos \chi_{k} - i \sin \chi_{k}
\hadj{\gamma}_{k} \hadj{\gamma}_{-k}) \ket{0_{k}}, \label{eq-state}
\end{equation}
where $\chi_{k} \equiv \theta_{k}^{\parent{2}} -
\theta_{k}^{\parent{1}}$, and $\ket{0_{k}}$ is the vacuum state
defined by $\gamma_{k} \ket{0_{k}} = \gamma_{-k} \ket{0_{k}} = 0$.
The time evolution then reads $\ket{\psi(t)} = \exp (-i H t)
\ket{\psi \parent{0}}$, and the LE takes the form
\begin{equation}
\mathcal{L} \parent{t} = \prod_{k>0} \parent{ 1 - A_k
\sinep{\Lambda_k t}{2} }, \label{eq-LE-2}
\end{equation}
where the coefficient $A_k \equiv \sinep{2 \chi_k}{2}$ is a slowly
varying positive function of the momentum $k$.

Now we show how the revivals in the LE can be derived from the
dispersion relation $\Lambda_k$ of the quasiparticles. We first take
the logarithm of Eq. (\ref{eq-LE-2}) to transform the product into a
sum over momentum $k$:
\begin{equation}
\ln \mathcal{L} \parent{t} = \sum_{k>0} \ln \parent{ 1 - A_k
\sinep{\Lambda_k t}{2} }. \label{eq-LE-3}
\end{equation}
Since each $k>0$ term has a periodicity $\pi / \Lambda_k$ in time,
and $A_k$ varies sufficiently slowly with $k$, this expression shows
that nearby $k$ modes separated by $\Delta k = 2\pi / N$ add up
constructively whenever $\Delta \Lambda_k t = p \pi$ with $p \in
\mathbb{Z}$. This rearranges to $t_k = \frac{1}{2} p N
\absval{\partial \Lambda_k / \partial k}^{-1}$, and in principle we
could expect a revival time $t_k$ corresponding to each $k$.
However, more modes can add up constructively if the dispersion
relation $\Lambda_k$ is closer to a straight line, therefore the
most pronounced revivals (in fact, the only revivals that stand out
from the background noise) are given by the stationary values of the
group velocity $v_g (k) \equiv \absval{\partial \Lambda_k /
\partial k}$. The first of these revivals is the one corresponding
to the maximal group velocity $v_{\mathrm{max}}=\max_k v_g (k)$ and
we can thus give the following estimate for the revival time scale:
\begin{equation}
T_{\mathrm{rev}} \simeq \frac{N} {2 \, v_{\mathrm{max}}} =
\frac{N}{2} \absval{\partder{\Lambda_k}{k}}_{\mathrm{max}}^{-1}.
\label{eq-rev}
\end{equation}
The maximum group velocity $v_{\mathrm{max}}$ can be computed
exactly from the dispersion relation $\Lambda_k$. It takes a
particularly simple form in the case of the Ising model ($\eta =
1$): $v_{\mathrm{max}} = h$ when $h < 1$ and $v_{\mathrm{max}} = 1$
when $h \geq 1$.

\begin{figure}[ht]
\centering
\subfigure[]{
\includegraphics[scale=0.3]{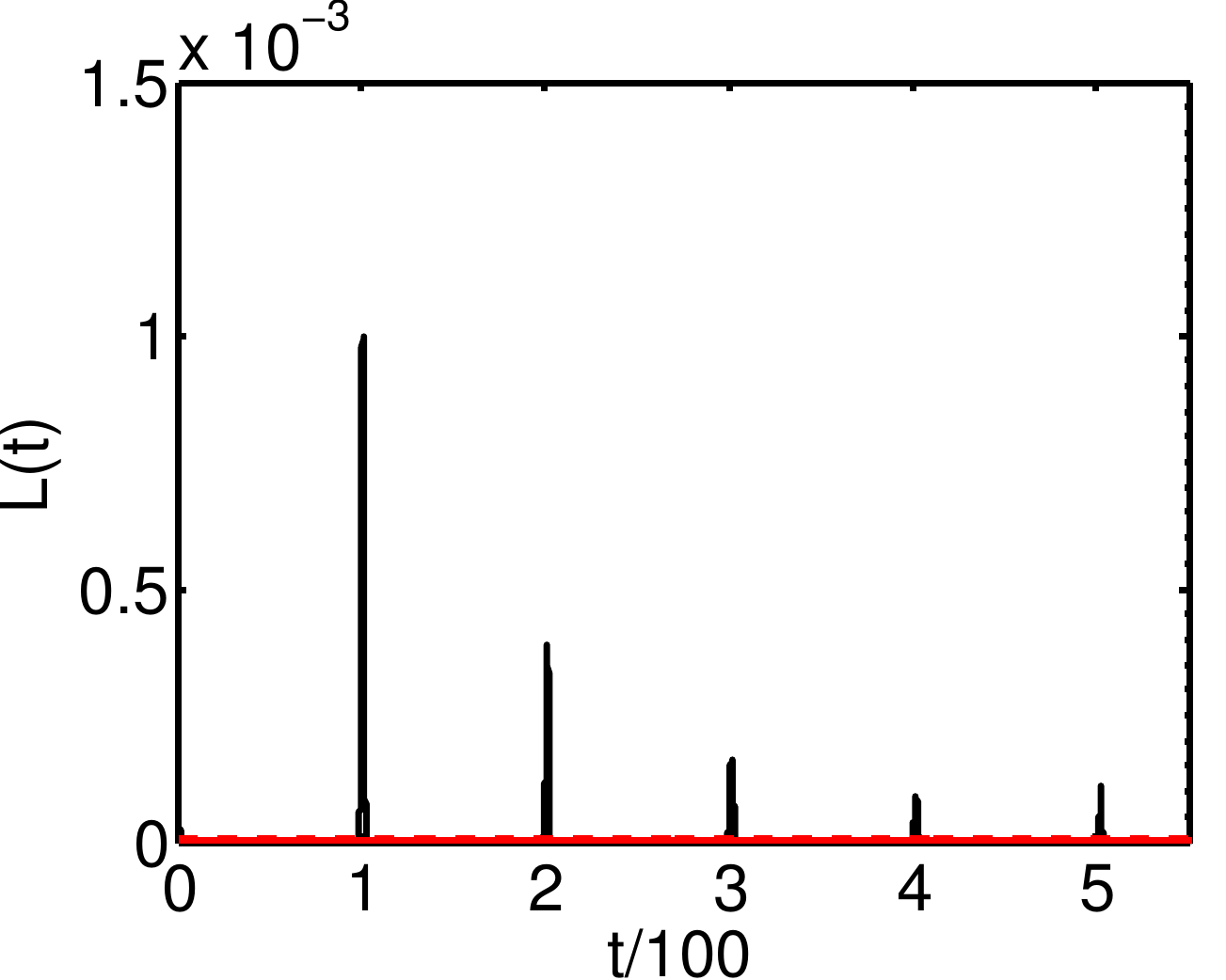}
\label{fig-1a} }
\subfigure[]{
\includegraphics[scale=0.3]{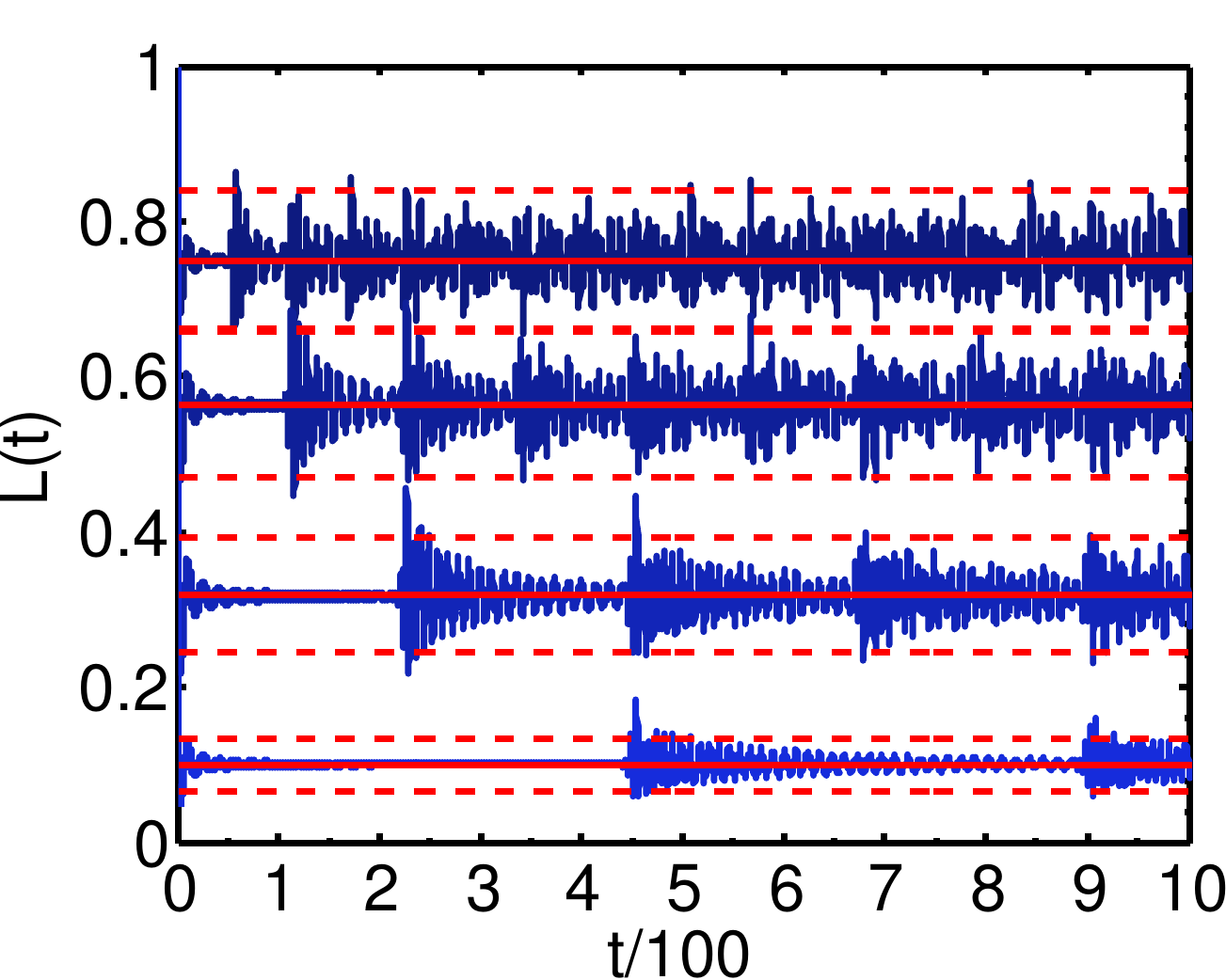}
\label{fig-1b}
}\\
\subfigure[]{
\includegraphics[scale=0.3]{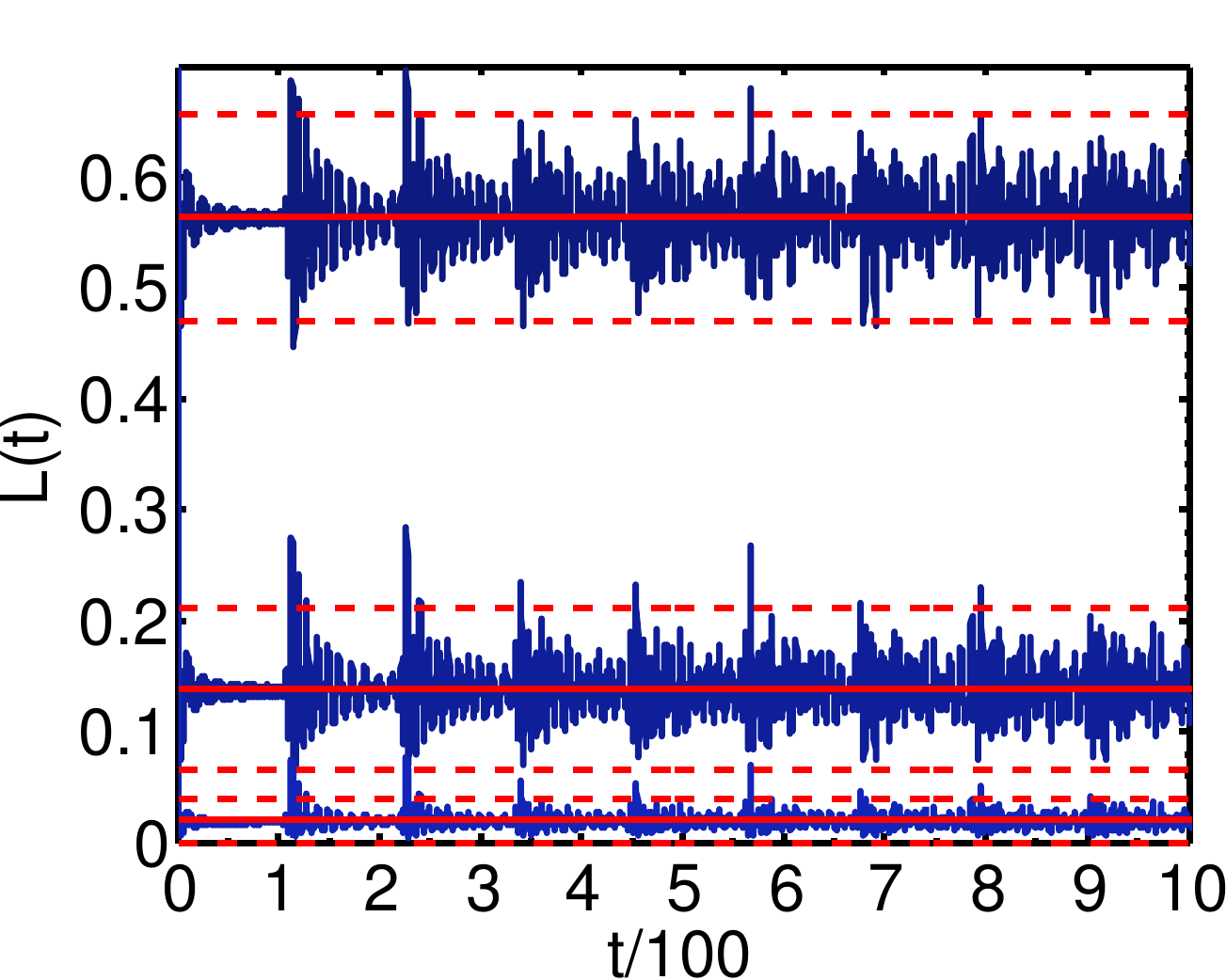}
\label{fig-1c} }
\subfigure[]{
\includegraphics[scale=0.3]{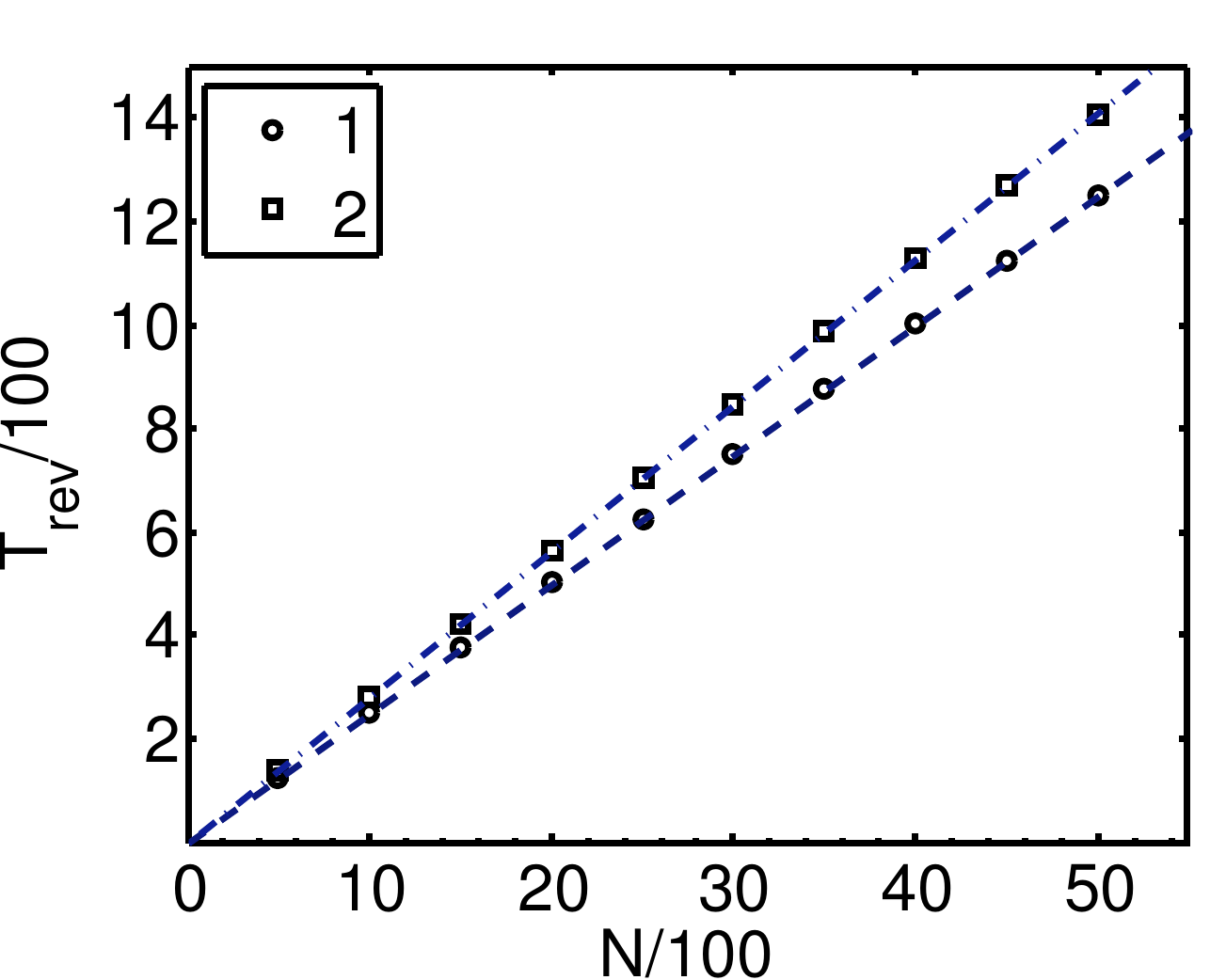}
\label{fig-1d} } \caption[] { Loschmidt echo for the quenched XY
model with antiperiodic boundary conditions ($q = 1$).
\subref{fig-1a} Critical quench with $\eta_1 = \eta_2 = 2.0$, $h_1 =
0.5$, $h_2 = 1.0$, $N = 400$, and $v_{\mathrm{max}} = 2.0$. Eq.
(\ref{eq-rev}) gives $T_{\mathrm{rev}}=100.0$, and from the plotted
data we get $T_{\mathrm{rev}} \approx 99.7$. \subref{fig-1b}
Non-critical quenches for different system sizes: $\eta_1 = \eta_2 =
2.0$, $h_1 = 0.7$, $h_2 = 0.8$, $N \in \sset{200, 400, 800, 1600}$,
and $v_{\mathrm{max}} \approx 1.77$. The equilibrium value of the LE
decreases with system size. Eq. (\ref{eq-rev}) gives
$T_{\mathrm{rev}} \approx 112.7$ for $N = 400$, and from the data we
get $T_{\mathrm{rev}} \approx 113.0$. \subref{fig-1c} Quenches of
different size: $\eta_1 = \eta_2 = 2.0$, $h_1 \in \sset{0.5, 0.6,
0.7}$, $h_2 = 0.8$, and $N=400$. Group velocities and revival times
match those in subfigure \subref{fig-1b}. \subref{fig-1d} Linear
scaling of the revival times with system size. The lines represent
the values given by Eq. (\ref{eq-rev}). Parameters used: 1: those in
subfigure \subref{fig-1a}, 2: those in subfigure \subref{fig-1b}. In
each plot, the red horizontal lines represent average values and
three standard deviations thereof. } \label{fig-1}
\end{figure}

In Fig. \ref{fig-1a}, the LE for a critical quench ($h_2=1$) with
antiperiodic boundary conditions ($q=1$) is plotted. The
$T_{\mathrm{rev}}$ predicted by Eq. (\ref{eq-rev}) is in perfect
agreement with the data. The same critical quench with periodic
boundary conditions ($q=0$) is interesting. At the odd revivals,
there is no signal in the LE due to a destructive (vanishing)
contribution from one of the $k$ modes in the product of Eq.
(\ref{eq-LE-2}). Only the even revivals are spotted.

Figs. \ref{fig-1b} and \ref{fig-1d} verify that the revival times
scale like $O(L)$ where $L \sim N$ is the linear size of the system.
The quench in Fig. \ref{fig-1b} is non-critical, and the parameters
of the quench Hamiltonian are far away from any phase boundaries.
This means that the system is gapped, and a simple spectral analysis
would imply that $T_{\mathrm{rev}}$ is of $O(1)$. However, even if
there is a small reconstruction of the wave-function at a time scale
$1 / \Delta$ (where $\Delta$ is a difference between any two energy
levels), the weight involved is not sufficiently large to make the
corresponding revival strong enough. \emph{Visible revivals} are
governed by Eq. (\ref{eq-rev}).

In Fig. \ref{fig-1c}, the system is quenched from different ground
states corresponding to different parameter values $\lambda^{(1)}$.
As one can see, the details of the evolution and the average values
$\overline{\mathcal L}$ are different, but the structure of the
revivals is the same for all the quenches, and the revival times are
consistent with those predicted by Eq. (\ref{eq-rev}). This is the
promised \emph{universality} of the revival structure: the initial
state and the size of the quench are unimportant. The parameters
$\lambda^{(2)}$ determine $v_{\mathrm{max}}$ and therefore
$T_{\mathrm{rev}}$ but not the fact that revivals happen at time
instances spaced evenly {at intervals
that are linear in system size}: $t_p = p L / v_{\mathrm{max}}$.

\subsection{Magnetization and entanglement entropy}

Although the behavior of the LE is illuminating for theoretical
arguments, it is hardly of experimental relevance because the
amplitude of the signal is zero in the thermodynamic limit. On the
other hand, it is expected that a revival in the LE corresponds to
revivals in macroscopic observables as well. To demonstrate this, we
study the time evolution of the order parameter (magnetization)
$\mu(t)$ \cite{igloi} and of the single spin entanglement entropy $S(t)$. If we
consider a subsystem $A$ consisting of a single spin, the von
Neumann entropy between subsystem $A$ and the rest of the system is
$S = -\trace{\rho_A \log{\rho_A}}$, where $\rho_A$ is the reduced
density matrix of subsystem $A$.

\begin{figure}[ht]
\centering
\includegraphics[scale=.4]{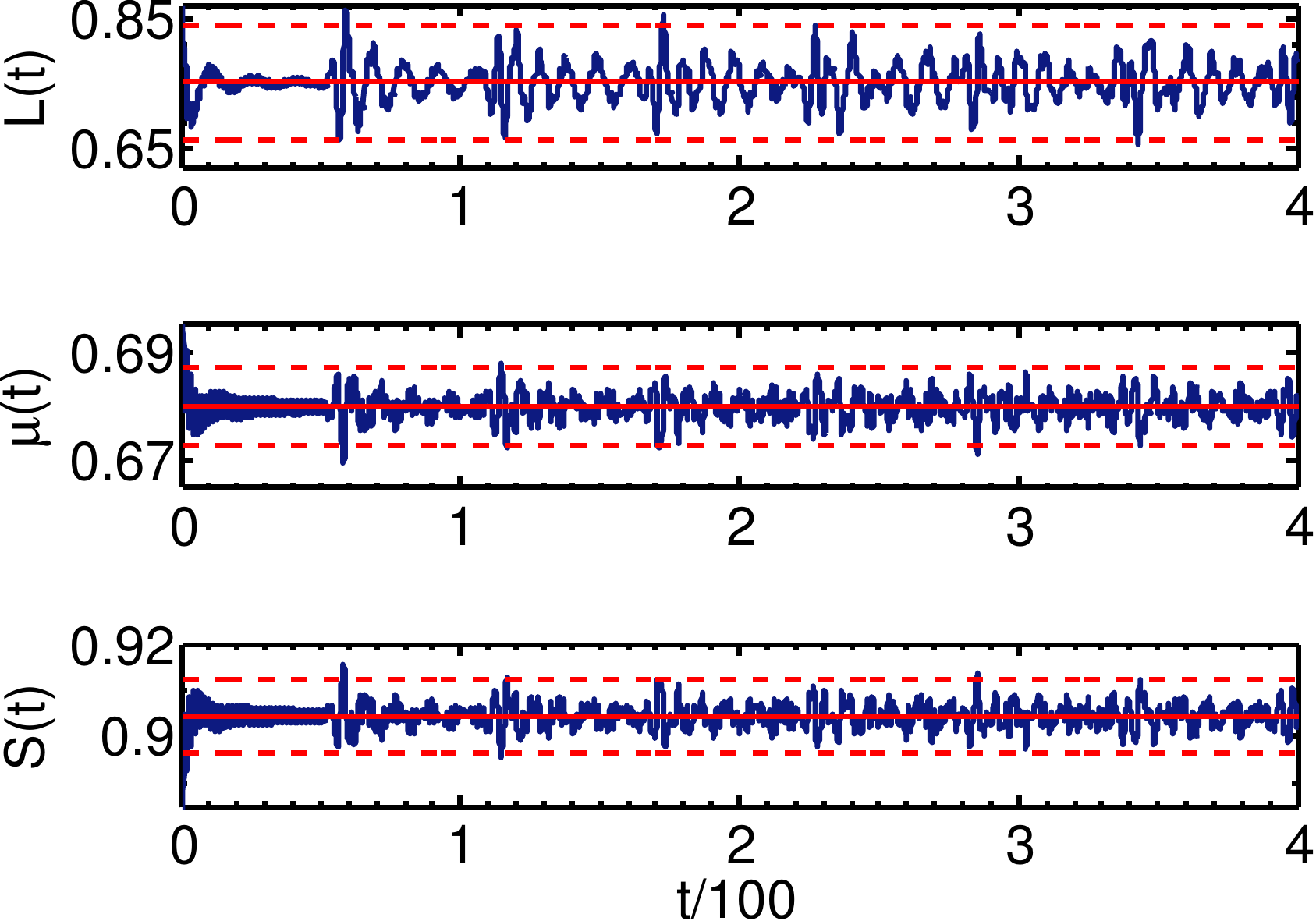}
\caption[] { A small quench with $q = 1$, $\eta_1 = \eta_2 = 2.0$,
$h_1 = 0.7$, $h_2 = 0.8$, $N = 200$, and $v_{\mathrm{max}} \approx
1.77$. From top to bottom: Loschmidt echo $\mathcal L (t)$,
magnetization $\mu (t)$, and single spin entanglement entropy $S
(t)$. Eq. (\ref{eq-rev}) gives $T_{\mathrm{rev}} \approx 56.37$, and
from the plotted data we get $T_{\mathrm{rev}} \approx 56.58$. The
horizontal lines represent average values and three standard
deviations thereof. } \label{fig-2}
\end{figure}

When the initial state of the quenching process in Eq.
(\ref{eq-state}) is expanded in position basis rather than momentum
basis, one finds that the excitations are pairwise correlated
between different lattice sites due to the fermionic anticommutation
relations. This implies that $\rho_A$ is diagonal, and then we can
use translational invariance to establish
\begin{equation}
S \parent{t} = - \mu \parent{t} \log{\mu \parent{t}} -
\parent{1 - \mu \parent{t}} \log{\parent{1 - \mu \parent{t}}}, \label{eq-S}
\end{equation}
where the average magnetization is given by
\begin{eqnarray}
\mu \parent{t} &=& \frac{1}{N} \sum_{l} \langle \hadj{c}_l c_l
\rangle = \frac{1}{N} \sum_{k} \langle \hadj{c}_k c_k
\rangle, \label{eq-mu} \\
\langle \hadj{c}_k c_k \rangle &=& \sinep{\theta_k}{2}
\cosinep{\chi_k}{2} + \cosinep{\theta_k}{2} \sinep{\chi_k}{2}
\nonumber \\
&-& 2 \, \sine{\theta_k} \cosine{\theta_k} \sine{\chi_k}
\cosine{\chi_k} \cosine{2 \Lambda_k t}. \nonumber
\end{eqnarray}
The entanglement entropy $S$ and the average magnetization $\mu$ are
thus governed by the interference of the same modes as in Eq.
(\ref{eq-LE-2}) with their frequencies given by the same dispersion
relation $\Lambda_k$. This means that they share the same time
scales as the LE. The results are illustrated in Fig. \ref{fig-2}.
In fact, since the magnetization is given as a sum rather than a
product of different oscillating modes it can be used instead of the
LE to determine $T_{\mathrm{rev}}$ in the large $N$ limit when the
LE has a very small average value. In practice, the magnetization is
analogous to the Logarithmic Loschmidt Echo used in
\cite{stephan-dubail}.

\subsection{Time evolution of a local disturbance}

\begin{figure}[ht]
\centering
\subfigure[]{
\includegraphics[scale=0.5]{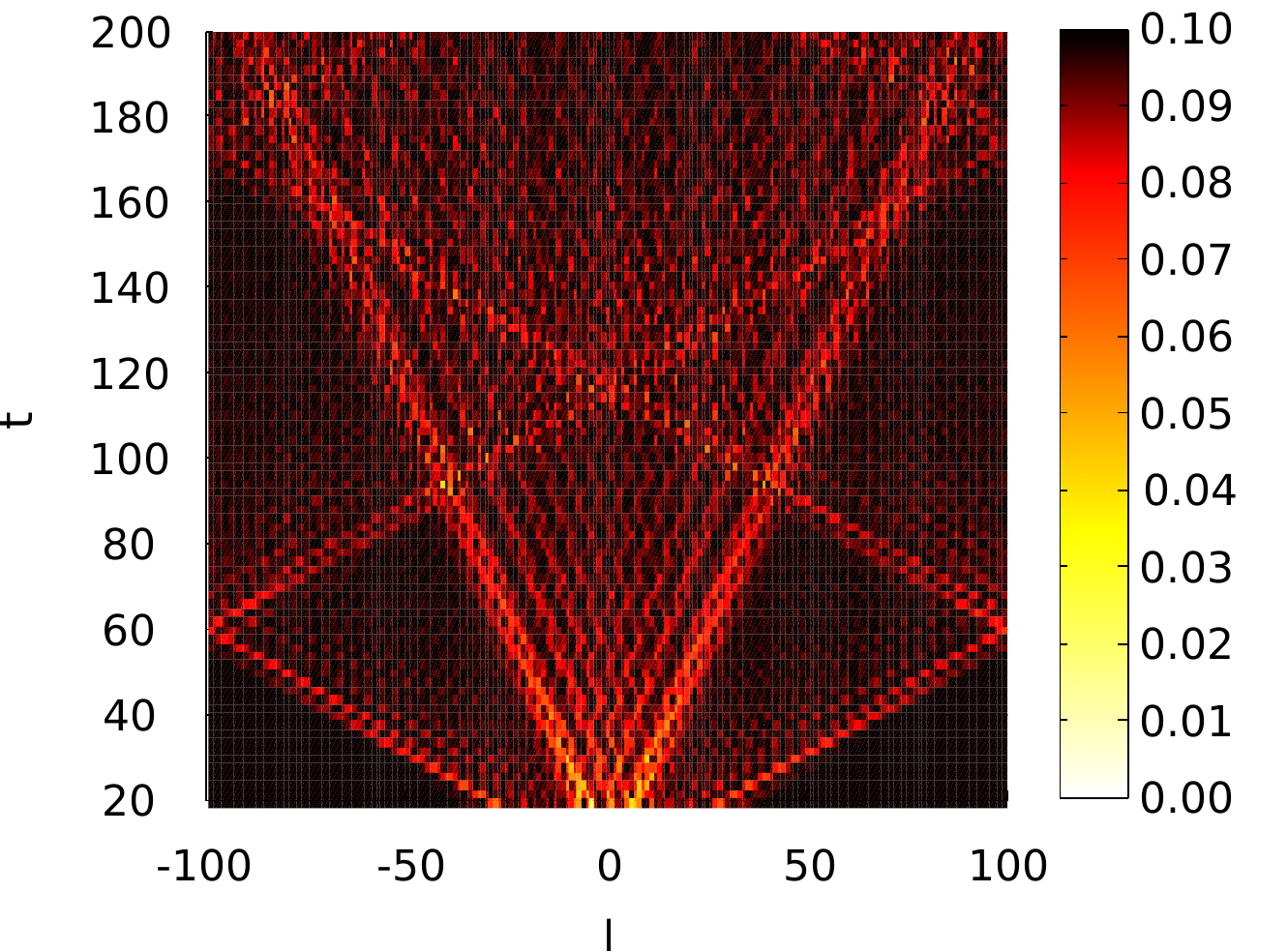}
\label{fig-3a} }
\subfigure[]{
\includegraphics[scale=0.5]{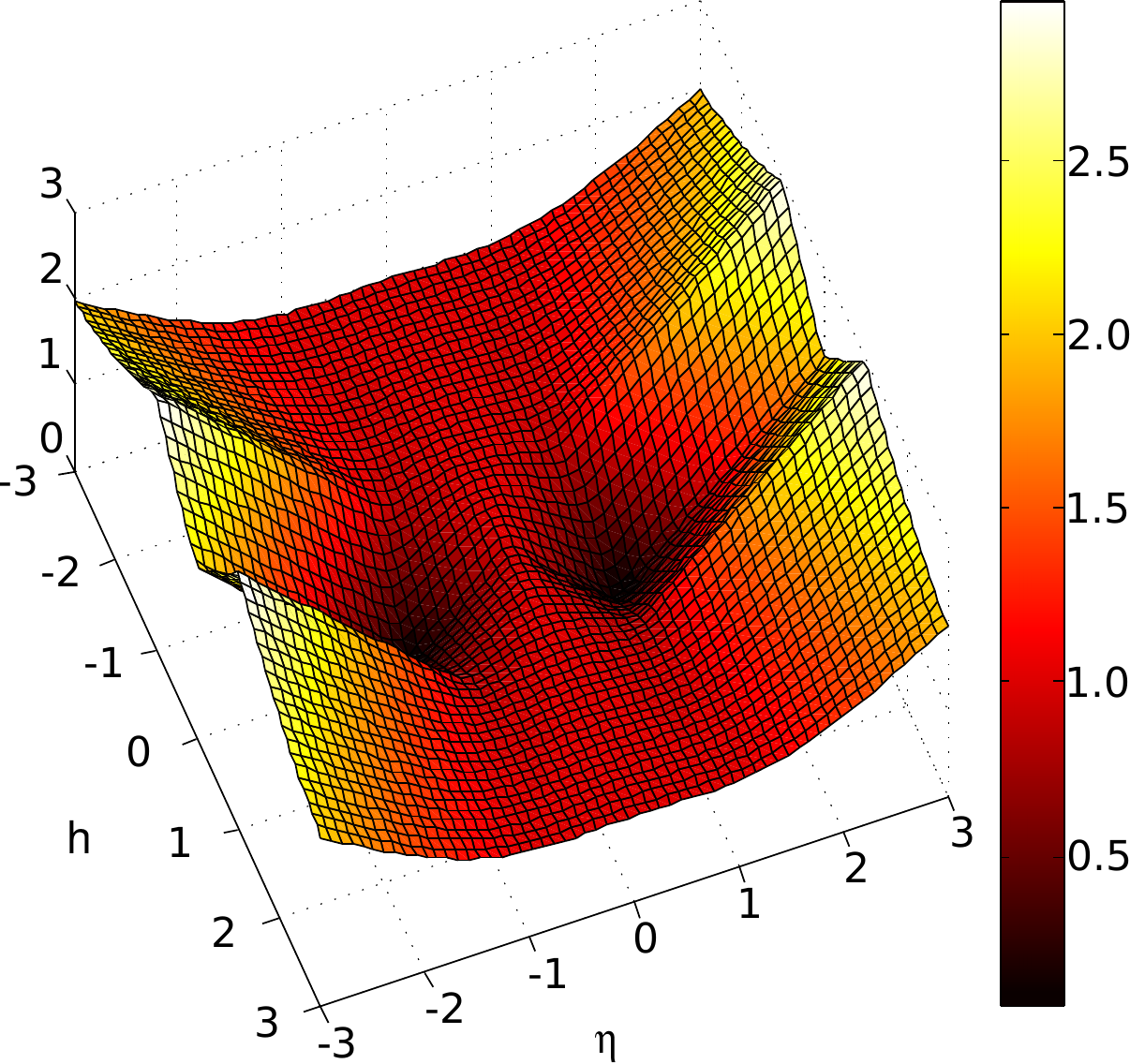}
\label{fig-3b} } \caption[] {\subref{fig-3a} Time evolution of a
local disturbance (local quench) $\absval{\Omega_l(t)}^2$ with $q =
1$, $\eta_1 = \eta_2 = 2.0$, $h_1 = 0.7$, $h_2 = 0.8$, and $N=200$.
The maximum speed is $v_{\mathrm{max}} \approx 1.77$. The
corresponding global quench is illustrated in Fig. \ref{fig-1b} for
comparison. \subref{fig-3b} Maximal group velocity
$v_{\mathrm{max}}$ as a function of $\eta$ and $h$. } \label{fig-3}
\end{figure}

Now we investigate the time evolution after a local disturbance in
the spin chain. In particular, we consider a single spin flip at
position $l$, which is represented by the operator $F_l = c_l +
\hadj{c}_l$. The time evolution of the resulting local disturbance
is best studied in the Heisenberg picture, where the operator $F_l$
becomes time dependent and takes the form
\begin{eqnarray}
F_l (t) &=& \sum_{l'} \left[ {\Omega_{l-l'} (t) c_{l'} +
\conj{\Omega}_{l-l'} (t) \hadj{c}_{l'}} \right], \label{eq-flip} \\
\Omega_{l-l'} (t) &=& \frac{1}{N} \sum_{k} e^{i k (l-l')} \Big{[}
e^{i \Lambda_k t} \sin^2 \theta_k + e^{-i \Lambda_k t} \cos^2
\theta_k \nonumber \\
&+& i \left( e^{-i \Lambda_k t} - e^{i \Lambda_k t} \right) \sin
\theta_k \cos\theta_k \Big{]}. \nonumber
\end{eqnarray}
This expression shows that nearby modes at momentum $k$ add up
constructively whenever $\Delta \Lambda_k t = 2p \pi \pm \Delta k |l
- l'|$ with $p \in \mathbb{Z}$. Due to $\Delta k = 2\pi / N$, this
condition can be written as $t = (p N \pm |l - l'|) \absval{\partial
\Lambda_k / \partial k}^{-1}$, and one verifies that the disturbance
travels with the group velocity $v_g (k) = \absval{\partial
\Lambda_k / \partial k}$. Once again, more such modes can add up
constructively if the second derivative of $\Lambda_k$ vanishes,
therefore we expect visible wave packets to travel with the
stationary values of $v_g (k)$. Indeed, Fig. \ref{fig-3a} shows that
the different wave packets corresponding to the local extrema of
$v_g (k)$ propagate through the lattice while maintaining their
respective wave forms.

The fastest wave packets travel with the maximal group velocity
$v_{\mathrm{max}}$, and the first revivals can be interpreted as
constructive interferences between them. Since we assume cyclic
boundary conditions, {one can think of the spin chain as a
closed ring}. In this {picture}, the fastest wave packets
first meet halfway in the ring at time $t = N / 2 v_{\textrm{max}}$
which indeed coincides with the first revival time in Eq.
(\ref{eq-rev}).

\section{Non-integrable perturbations} \label{sec-pert}

{In the previous section, we showed that} the structure of
{the} revivals is governed by the maximal speed of
{quasiparticles} in the system. We {found} that as
long as the {quasiparticles} exist, the details of the quench
are not relevant. {There is} universality within the
integrable behavior of the system. At this point, we wonder whether
there is universality beyond the quasi-free system. {The
local physics} induced by the local Hamiltonian might imply that as
long as information is not completely lost $-$ as in the case of an
infinite system $-$ {revivals} can be {observed due to
the} recombination of the fastest signals even when these are not
point-like and do not correspond to {quasiparticles}.
{To investigate this possibility, we now study} the
robustness of the revival structure against non-integrable
perturbations.

\subsection{XZ spin chain}

We start by considering the XZ spin chain, which contains an
additional $\sigma_l^z \sigma_{l+1}^z$ coupling with respect to the
quantum Ising model ($\eta = 1$):
\begin{equation}
H = - \frac{1}{2}\sum_{l=1}^{N} \parent{ \sigma_l^x \sigma_{l+1}^x +
g \sigma_l^z \sigma_{l+1}^z + h \sigma_l^z }. \label{eq-H-3}
\end{equation}
This Hamiltonian is non-integrable, and we simulate the quantum
quench by exact numerical diagonalization. To achieve a relatively
large system size ($N \simeq 50$), we take the limit of large field
($h \gg 1$), restricting the effective Hilbert space to states where
almost all spins are aligned with the field.

\begin{figure}[h]
\centering
\subfigure[]{
\includegraphics[scale=.2]{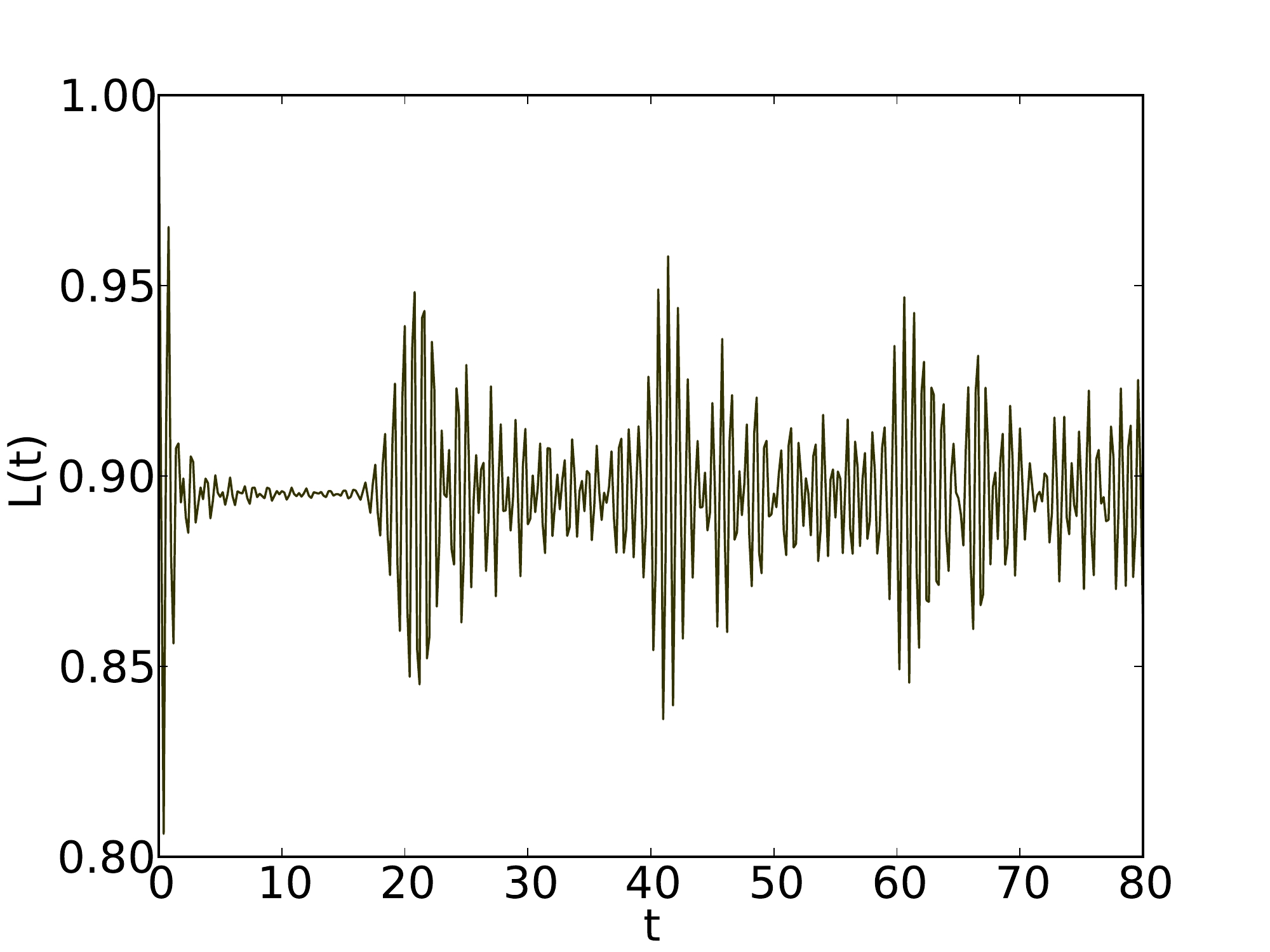}
\label{fig-4a} } \subfigure[]{
\includegraphics[scale=.2]{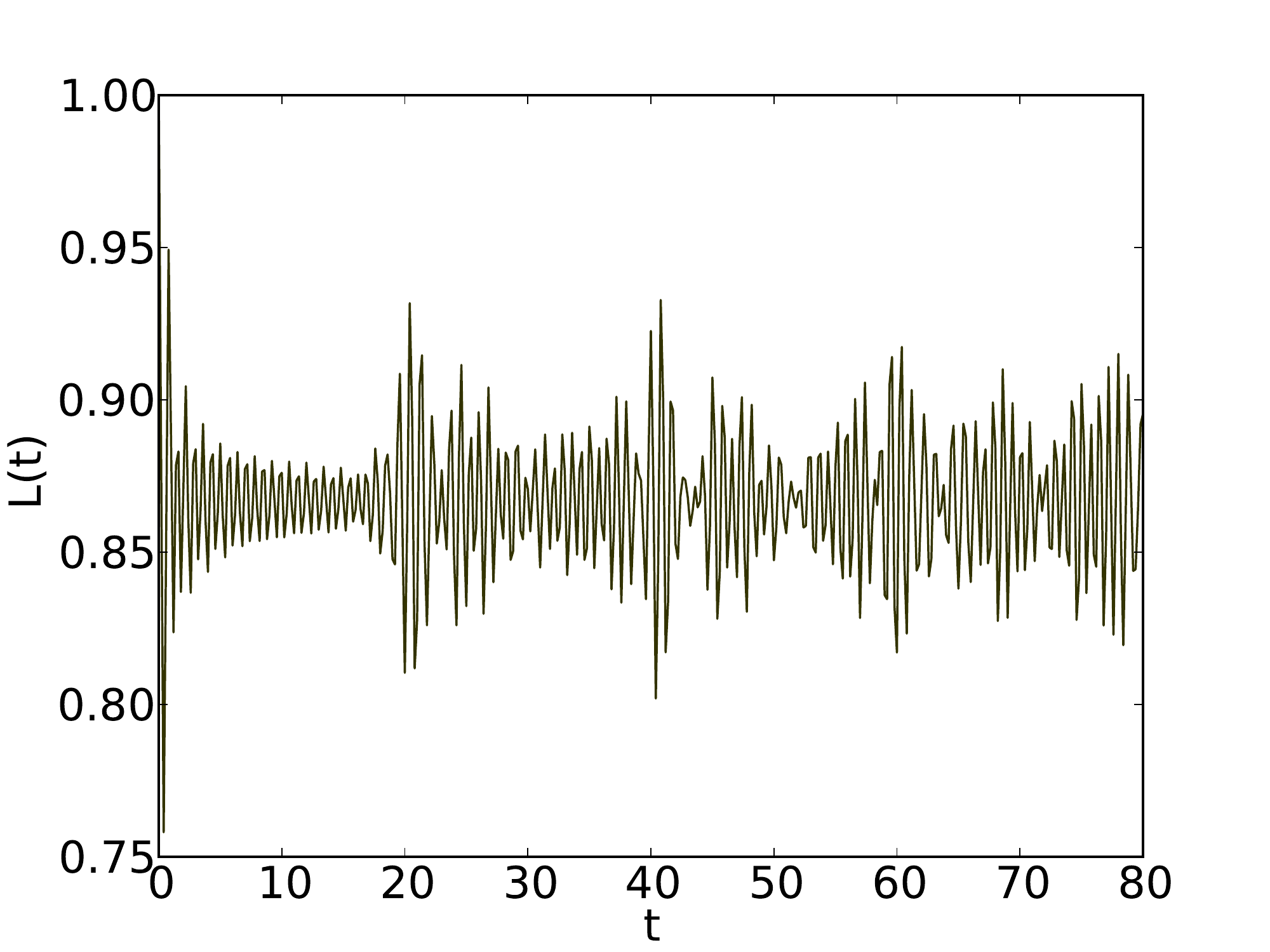}
\label{fig-4b} }
\caption[] { Loschmidt echo for the quenched XZ spin chain with
periodic boundary conditions in the integrable case $g_2 = 0$ (a)
and the non-integrable case $g_2 = 0.3$ (b). The other quench
parameters are $\eta_1 = \eta_2 = 1.0$, $h_1 = 10.0$, $h_2 = 4.0$,
and $g_1 = 0.0$ in both subfigures. Since $N = 40$ and
$v_{\mathrm{max}} = 1$, Eq. (\ref{eq-rev}) gives $T_{\mathrm{rev}} =
20$.
\label{fig4}
}
\end{figure}


As shown in Fig. \ref{fig4}, the structure of the revivals
is clearly visible for $|g| \lesssim 0.5$, and hence this structure
is universal for a \emph{range} of the non-integrability parameter
$g$. The revival times scale linearly with $N$, and the range of
visibility is largely independent of both the system size $N$ and
the magnetic field $h$. We find that pronounced revivals
{gradually} disappear {in the range} $0.3 < |g| < 0.7$
for all $20 \leq N \leq 50$ when $h = 4$ is fixed, and for all $2
\leq h \leq 100$ when $N = 20$ is fixed. Since $v_{\mathrm{max}}
\sim 1$ in the Ising model for $h \geq 1$, we see that the
non-equilibrium dynamics is dominated by the $\sigma_l^x
\sigma_{l+1}^x$ term in Eq. (\ref{eq-H-3}). The XZ spin chain is
therefore significantly non-integrable for $|g| \sim 0.5$, even in
the $h \rightarrow \infty$ limit. This claim is further supported by
the fact that the visibility range in $g$ does not depend on $h$. We
finally note that the equilibrium (long-term average) value of the
magnetization is strongly dependent on the initial state and so this
equilibration is not thermalization, even though the system is
non-integrable \cite{mueller}.

\subsection{Random disorder in the field}

Now we consider another integrability breaking perturbation to the
XY spin chain. We introduce a site dependent external field
component $e_l$ to the Hamiltonian that explicitly breaks the
translational invariance and the integrability of the model.
{The total site dependent field is} $h_l = h + e_l$, where
$e_l$ is randomly picked from a uniform distribution with a maximum
amplitude $\epsilon$ in the sense $-\epsilon < e_l < \epsilon$.

\begin{figure}[h]
\centering
\subfigure[]{
\includegraphics[scale=0.3]{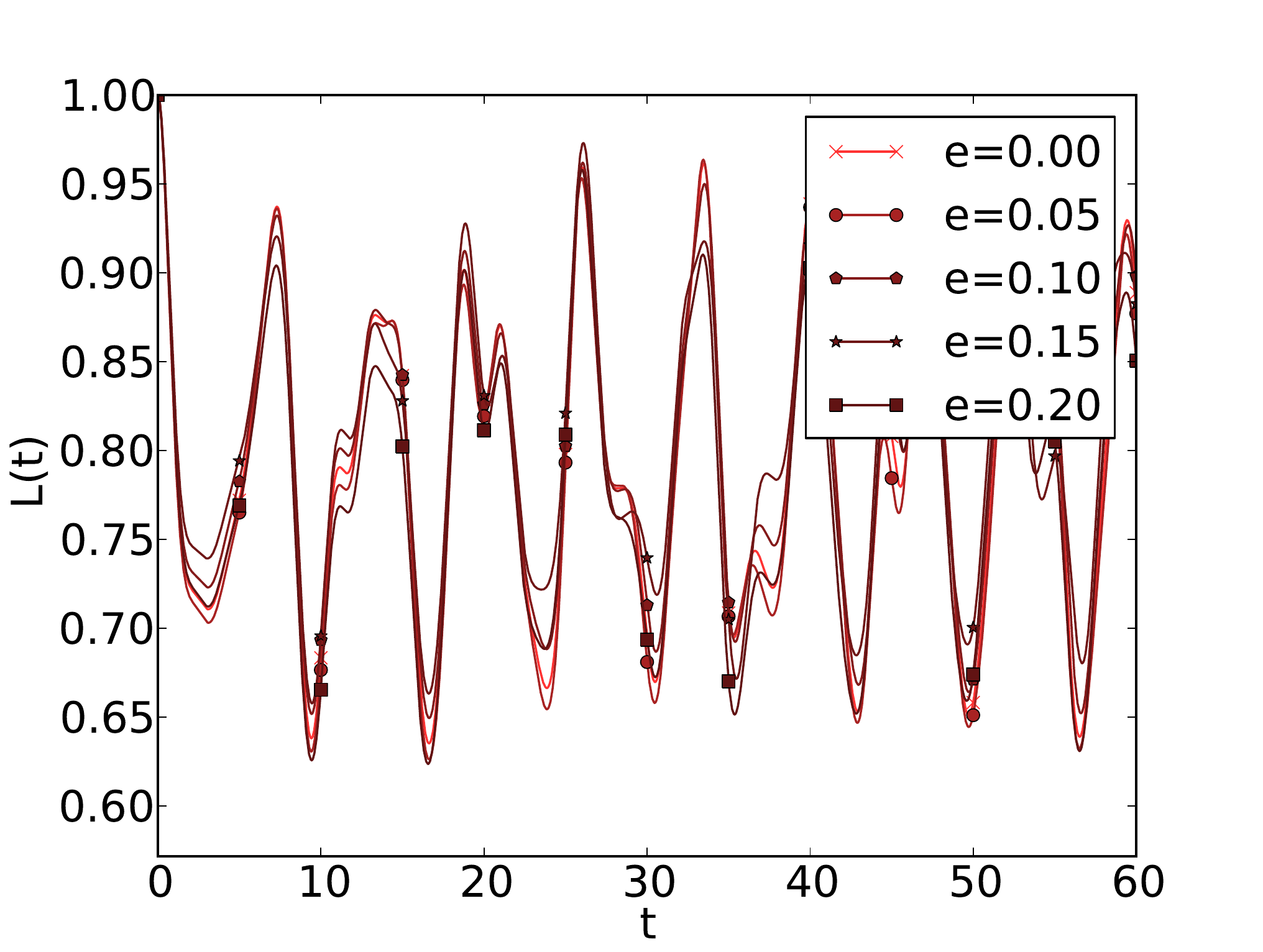}
\label{fig-5a} } \subfigure[]{
\includegraphics[scale=0.3]{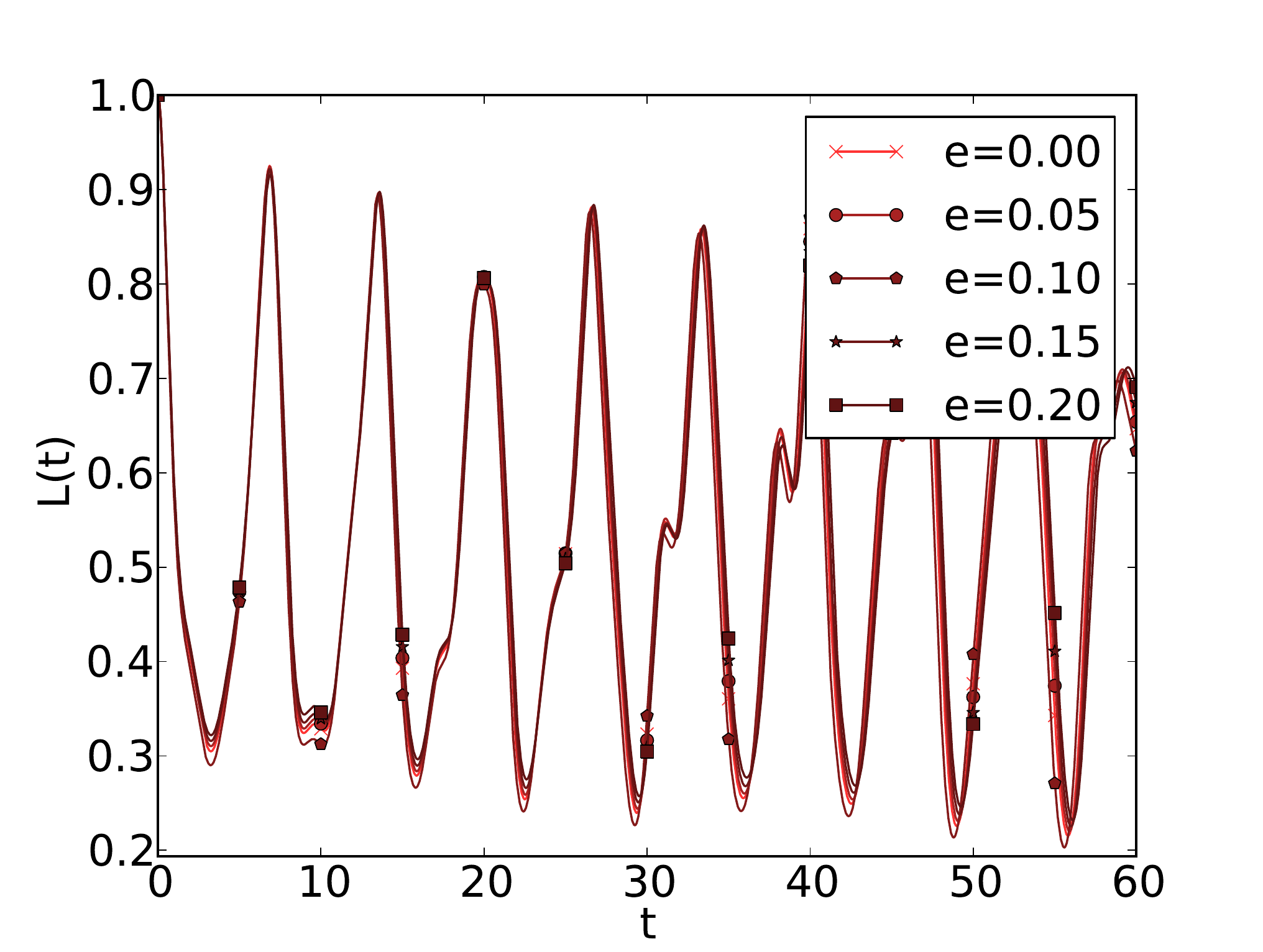}
\label{fig-5b} } \caption[]{\label{fig-5} Loschmidt echo for the
transverse field Ising model when the external field has a small
site dependent component of varying magnitude: $\eta_1 = \eta_2 =
1.0$, $N = 13$, periodic boundary conditions. \subref{fig-5a}
Non-critical case: $h_1 = 0.5$, $h_2 = 0.8$, $\epsilon_1 = 0.0$,
$\epsilon_2 = e$. Eq. \eqref{eq-rev} gives $T_{\mathrm{rev}}
= 8.12$ for $\epsilon = 0.0$. \subref{fig-5b} Critical case: $h_1 =
0.5$, $h_2 = 1.0$, $\epsilon_1 = 0.0$, $\epsilon_2 = e$. Eq.
\eqref{eq-rev} gives $T_{\mathrm{rev}} = 6.5$ for $\epsilon = 0.0$.
}
\end{figure}

Using exact diagonalization and exploiting the fact that the
Hamiltonian decomposes to subspaces of odd and even number of spins,
we can assess the effect of the site dependent field disturbance for
modest sized systems $N \leq 13$. The simulations run on such
systems {give supporting evidence that the revival structure
is essentially unchanged for a range of the amplitude $\epsilon$}.
In Fig. \ref{fig-5}, the Loschmidt echo for a critical and a
non-critical system is plotted {with} various values of
$\epsilon$ to illustrate {this observation}.

\section{Discussion} \label{sec-disc}

We found in Section \ref{sec-xy} that the transverse field XY
model exhibits a universal structure of revivals that is independent
of the initial state and the size of the quench. Since the XY model
is integrable and its exact solution is in terms of free fermionic
quasiparticles, there is a straightforward interpretation for the
phenomenon of revivals. The information propagates around the system
via wave packets of quasiparticles, and the first revival occurs
when the wave packets traveling with the maximal group velocity
$v_{\mathrm{max}}$ meet. This interpretation explains the
universality of the revival structure since $v_{\mathrm{max}}$ only
depends on the dispersion relation of the quasiparticles associated
with the quench Hamiltonian (and nothing which would be related to
the initial state).

On the other hand, the robustness against non-integrable
perturbations found in Section \ref{sec-pert} suggests something
more generic than the quasiparticle interpretation. The
quasiparticles of the perturbed system are no longer free but there
is a finite interaction between them. For the XZ spin chain, this
interaction is on the order of $g$ as can be verified by looking at
the exact eigenvalues for finite systems. This implies that the
quasiparticles decay on the time scale of $1/g$, and hence one would
not expect to see revivals at time scales {much} larger than
$1/g$. However, the evidence is on the contrary: at $N = 50$ and $g
\sim 0.5$, revivals are still clearly visible at $t \sim 50$, which
is an order of magnitude larger than $1/g \sim 2$.

It appears that the revival structure is a non-equilibrium property
that is beyond integrability and the existence of stable
quasiparticles. Here we provide a more generic interpretation in
terms of \emph{locality}. Quantum many-body physics is generally
described by Hamiltonians that can be written as sums of local
operators. The locality of the Hamiltonian has profound consequences
on the dynamics of the system \cite{lieb-robinson}: there exists an
emergent light cone for the propagation of information such that
signals outside the light cone are exponentially suppressed. The
characteristic speed of the light cone gives the maximal speed of
information in the system, and it is called the Lieb$-$Robinson
speed $v_{LR}$. In general, it depends on both the graph of the
system and the strengths of the interactions in the Hamiltonian.

We speculate that the revival structure described in the previous
sections is much more generic and valid whenever a many-body quantum
system has local dynamics, as long as the {integrability
breaking} is {not too strong and hence} information is
{not} completely lost in the system. Since the propagation of
information is governed by $v_{LR}$, the revival time scale becomes
$T_{\mathrm{rev}} \simeq L / v_{LR}$ in general. In the integrable
case, $v_{LR}$ reduces to the quasiparticle speed
$v_{\mathrm{max}}$, and hence we recover the revival time scale in
Eq. (\ref{eq-rev}). On the other hand, the locality of the dynamics
is intact in the non-integrable case as well, providing a
natural explanation for the robustness of the revival structure. \\

\section{Summary} \label{sec-summ}

In this paper, we studied the phenomenon of revivals after a quantum
quench in the transverse field XY model, and found a non-trivial
revival structure that can not be obtained from a simple spectral
analysis. It was shown that this structure is universal in the sense
that it does not depend on the initial state and the size of the
quench. Revivals were shown to be related to quasiparticles
propagating around the system with a finite maximum speed
$v_{\mathrm{max}}$, and a corresponding estimate for the revival
time scale was established.

We also investigated the effect of non-integrable perturbations on
the structure of the revivals. In particular, we considered the XZ
spin chain and a random disorder in the magnetic field. It was found
that the revival structure is clearly visible at time scales far
beyond the lifetime of quasiparticles, implying that something more
generic than integrability is behind the phenomenon of revivals. In
perspective of this, we proposed a generic connection between
revivals and locality, where the quasiparticle speed
$v_{\mathrm{max}}$ generalizes into the Lieb$-$Robinson speed
$v_{LR}$. We believe that a thorough understanding of this important
connection requires further study of non-integrable systems, for
example, studying how the entanglement production in the subsystem
is related to the loss of {the revival} structure.

\begin{acknowledgments}

We thank Lorenzo Campos Venuti and Paolo Zanardi for important
discussions. Research at Perimeter Institute for Theoretical Physics
is supported in part by the Government of Canada through NSERC and
by the Province of Ontario through MRI.

\end{acknowledgments}


\begin{thebibliography}{31}
\expandafter\ifx\csname natexlab\endcsname\relax\def\natexlab#1{#1}\fi
\expandafter\ifx\csname bibnamefont\endcsname\relax
  \def\bibnamefont#1{#1}\fi
\expandafter\ifx\csname bibfnamefont\endcsname\relax
  \def\bibfnamefont#1{#1}\fi
\expandafter\ifx\csname citenamefont\endcsname\relax
  \def\citenamefont#1{#1}\fi
\expandafter\ifx\csname url\endcsname\relax
  \def\url#1{\texttt{#1}}\fi
\expandafter\ifx\csname urlprefix\endcsname\relax\def\urlprefix{URL }\fi
\providecommand{\bibinfo}[2]{#2}
\providecommand{\eprint}[2][]{\url{#2}}

\bibitem{coldatoms} M. Greiner, O. Mandel, T. Esslinger, T. W. H\"{a}nsch, and I. Bloch,
Nature \textbf{415}, 39 (2002); M. Greiner, O. Mandel, T. W.
H\"{a}nsch, and I. Bloch, Nature \textbf{419}, 51 (2002).

\bibitem{review} A. Polkovnikov, K. Sengupta, A. Silva, and M. Vengalattore, Rev. Mod.
Phys. \textbf{83}, 863 (2011).

\bibitem{kibble-zurek} T. Kibble, Physics Today \textbf{60}, 47 (2007); W. H. Zurek, Phys. Rep. \textbf{276}, 177 (1996).

\bibitem[{\citenamefont{Tasaki}(1998)}]{tasaki}
\bibinfo{author}{\bibfnamefont{H.}~\bibnamefont{Tasaki}},
\bibinfo{journal}{\PRL} \textbf{\bibinfo{volume}{80}}, \bibinfo{pages}{1373} (\bibinfo{year}{1998});
\bibinfo{author}{\bibfnamefont{S.}~\bibnamefont{Goldstein}},
\bibinfo{author}{\bibfnamefont{J.}~\bibfnamefont{L.}~\bibnamefont{Lebowitz}},
\bibinfo{author}{\bibfnamefont{R.}~\bibnamefont{Tumulka}}, \bibnamefont{and}
\bibinfo{author}{\bibfnamefont{N.}~\bibnamefont{Zangh\`{i}}},
\bibinfo{journal}{\PRL} \textbf{\bibinfo{volume}{96}},
\bibinfo{pages}{050403} (\bibinfo{year}{2006}); \bibinfo{author}{\bibfnamefont{P.}~\bibnamefont{Reimann}},
\bibinfo{journal}{\PRL} \textbf{\bibinfo{volume}{99}},
\bibinfo{pages}{160404} (\bibinfo{year}{2007}); \bibinfo{author}{\bibfnamefont{P.}~\bibnamefont{Reimann}},
\bibinfo{journal}{\PRL} \textbf{\bibinfo{volume}{101}},
\bibinfo{pages}{190403} (\bibinfo{year}{2008}).

\bibitem[{\citenamefont{Popescu et~al.}(2006)\citenamefont{Popescu, Short, and Winter}}]{winter}
\bibinfo{author}{\bibfnamefont{S.}~\bibnamefont{Popescu}},
\bibinfo{author}{\bibfnamefont{A.~J.} \bibnamefont{Short}}, \bibnamefont{and}
\bibinfo{author}{\bibfnamefont{A.}~\bibnamefont{Winter}},
\bibinfo{journal}{Nature Physics} \textbf{\bibinfo{volume}{2}},
\bibinfo{pages}{754} (\bibinfo{year}{2006}); \bibinfo{author}{\bibfnamefont{N.}~\bibnamefont{Linden}},
\bibinfo{author}{\bibfnamefont{S.}~\bibnamefont{Popescu}},
\bibinfo{author}{\bibfnamefont{A.~J.} \bibnamefont{Short}}, \bibnamefont{and}
\bibinfo{author}{\bibfnamefont{A.}~\bibnamefont{Winter}},
\bibinfo{note}{arXiv:0812.2385}.

\bibitem{polkovnikov} A. Polkovnikov, Phys. Rev. B \textbf{72}, 161201(R) (2005); L. Campos Venuti and P. Zanardi, Phys. Rev. Lett. \textbf{99}, 095701
(2007).

\bibitem[{\citenamefont{Silva}(2008)}]{quench1}
\bibinfo{author}{\bibfnamefont{A.}~\bibnamefont{Silva}},
\bibinfo{journal}{\PRL} \textbf{\bibinfo{volume}{101}},
\bibinfo{pages}{120603} (\bibinfo{year}{2008}); \bibinfo{author}{\bibfnamefont{M.~A.} \bibnamefont{Cazalilla}},
\bibinfo{journal}{\PRL} \textbf{\bibinfo{volume}{97}},
\bibinfo{pages}{156403} (\bibinfo{year}{2006}).

\bibitem[{\citenamefont{Calabrese}(2006)}]{quench2}
\bibinfo{author}{\bibfnamefont{P.}~\bibnamefont{Calabrese}} \bibnamefont{and}
\bibinfo{author}{\bibfnamefont{J.}~\bibnamefont{Cardy}},
\bibinfo{journal}{\PRL} \textbf{\bibinfo{volume}{96}},
\bibinfo{pages}{136801} (\bibinfo{year}{2006}).

\bibitem{topquench} D. I. Tsomokos, A. Hamma, W. Zhang, S. Haas, and R. Fazio, Phys. Rev. A \textbf{80}, 060302(R)
(2009).

\bibitem{rigol1} M. Rigol, V. Dunjko, and M. Olshanii, Nature \textbf{452}, 854
(2008).

\bibitem{cramer1} M. Cramer, C. M. Dawson, J. Eisert, and T. J. Osborne, Phys. Rev. Lett. \textbf{100}, 030602 (2008); M. Cramer and J. Eisert, New J. Phys. \textbf{12}, 055020
(2010).

\bibitem[{\citenamefont{Cramer et~al.}(2008)\citenamefont{Cramer, Flesch, McCulloch, Schollw\"{o}ck, and Eisert}}]{cramer2}
\bibinfo{author}{\bibfnamefont{M.}~\bibnamefont{Cramer}},
\bibinfo{author}{\bibfnamefont{A.}~\bibnamefont{Flesch}},
\bibinfo{author}{\bibfnamefont{I.~P.} \bibnamefont{McCulloch}},
\bibinfo{author}{\bibfnamefont{U.}~\bibnamefont{Schollw\"{o}ck}},
\bibnamefont{and} \bibinfo{author}{\bibfnamefont{J.}~\bibnamefont{Eisert}},
\bibinfo{journal}{\PRL} \textbf{\bibinfo{volume}{101}},
\bibinfo{pages}{063001} (\bibinfo{year}{2008}).

\bibitem{rigol2} M. Rigol, V. Dunjko, V. Yurovsky, and M. Olshanii, Phys. Rev. Lett. \textbf{98}, 050405 (2007); M. A. Cazalilla, Phys. Rev. Lett. \textbf{97}, 156403 (2006).

\bibitem{rigol3} M. Rigol, Phys. Rev. A \textbf{80}, 053607 (2009).

\bibitem{mueller} C. Gogolin, M. P. M\"{u}ller, and J. Eisert, Phys. Rev. Lett. \textbf{106}, 040401
(2011).

\bibitem[{\citenamefont{Quan et~al.}(2006{\natexlab{a}})\citenamefont{Quan, Song, Liu, Zanardi, and Sun}}]{quan}
\bibinfo{author}{\bibfnamefont{H.~T.} \bibnamefont{Quan}},
\bibinfo{author}{\bibfnamefont{Z.}~\bibnamefont{Song}},
\bibinfo{author}{\bibfnamefont{X.~F.} \bibnamefont{Liu}},
\bibinfo{author}{\bibfnamefont{P.}~\bibnamefont{Zanardi}}, \bibnamefont{and}
\bibinfo{author}{\bibfnamefont{C.~P.} \bibnamefont{Sun}},
\bibinfo{journal}{\PRL} \textbf{\bibinfo{volume}{96}},
\bibinfo{pages}{140604} (\bibinfo{year}{2006}{\natexlab{a}}).

\bibitem{venuti} L. Campos Venuti and P. Zanardi, Phys. Rev. A \textbf{81}, 022113
(2010).

\bibitem[{\citenamefont{Barouch et~al.}(1970)\citenamefont{Barouch, McCoy, and Dresden}}]{barouch}
\bibinfo{author}{\bibfnamefont{E.}~\bibnamefont{Barouch}},
\bibinfo{author}{\bibfnamefont{B.~M.} \bibnamefont{McCoy}}, \bibnamefont{and}
\bibinfo{author}{\bibfnamefont{M.}~\bibnamefont{Dresden}},
\bibinfo{journal}{\PRA} \textbf{\bibinfo{volume}{2}}, \bibinfo{pages}{1075} (\bibinfo{year}{1970}).

\bibitem{stephan-dubail} J.-M. St\'{e}phan and J. Dubail,
arXiv:1105.4846.

\bibitem{lieb-robinson} E. H. Lieb and D. W. Robinson, Comm. Math. Phys. \textbf{28},
251 (1972);  B. Nachtergaele, Y. Ogata, and R. Sims, J. Stat. Phys.
\textbf{124}, 1 (2006).
\bibitem{igloi} Igloi and Rieger, Phys. Rev. Lett. 106, 035701 (2011)
\end{thebibliography}
\end{document}